# Meta-analysis using flexible random-effects distribution models

Hisashi Noma, PhD*
Department of Data Science, The Institute of Statistical Mathematics, Tokyo, Japan
ORCID: http://orcid.org/0000-0002-2520-9949

Kengo Nagashima, PhD
Research Center for Medical and Health Data Science, The Institute of Statistical Mathematics, Tokyo, Japan

Shogo Kato, PhD
Department of Statistical Inference and Mathematics, The Institute of Statistical Mathematics, Tokyo, Japan

Satoshi Teramukai, PhD
Department of Biostatistics, Graduate School of Medical Science, Kyoto Prefectural University of Medicine, Kyoto, Japan

Toshi A. Furukawa, MD, PhD
Departments of Health Promotion and Human Behavior, Kyoto University Graduate School of Medicine/School of Public Health, Kyoto, Japan

*Corresponding author: Hisashi Noma
Department of Data Science, The Institute of Statistical Mathematics
10-3 Midori-cho, Tachikawa, Tokyo 190-8562, Japan
TEL: +81-50-5533-8440, FAX: +81-42-526-4339
e-mail: noma@ism.ac.jp

## Abstract

**Background:** In meta-analysis, the normal distribution assumption has been adopted in most systematic reviews of random-effects distribution models due to its computational and conceptual simplicity. However, this restrictive model assumption is possibly unsuitable and might have serious influences in practices.

**Methods:** We provide two examples of real-world evidence that clearly show that the normal distribution assumption is explicitly unsuitable. We propose new random-effects meta-analysis methods using five flexible random-effects distribution models that can flexibly regulate skewness, kurtosis and tailweight: skew normal distribution, skew *t*-distribution, asymmetric Subbotin distribution, Jones–Faddy distribution, and sinh–arcsinh distribution. We also developed a statistical package, **flexmeta,** that can easily perform these methods.

**Results:** Using the flexible random-effects distribution models, the results of the two meta-analyses were markedly altered, potentially influencing the overall conclusions of these systematic reviews.

**Conclusions:** The restrictive normal distribution assumption in the random-effects model can yield misleading conclusions. The proposed flexible methods can provide more precise conclusions in systematic reviews.

## Introduction

In meta-analysis in medical studies, random-effects models have been the primary statistical tools for quantitative evaluation of treatment effects that account for between-studies heterogeneity[1, 2]. Conventionally, the normal distribution assumption has been adopted in most systematic reviews due to its computational and conceptual simplicity[2, 3]. However, the shape of the random-effects distribution reflects how the treatment effects parameters (e.g., mean difference, log relative risk) are distributed in the target population, and are directly associated with the fundamental heterogeneity of treatment effects. If the normal distribution assumption diverges drastically from the true heterogeneous structure, the overall results of the meta-analyses may be misleading. In addition, in recent studies, prediction intervals have been gaining prominence in meta-analyses as a means to quantify heterogeneity and effectiveness in real-world uses of the treatment [4, 5]. Because the prediction interval is constructed by the estimated random-effects distribution, it should be directly influenced by the form of the distribution assumptions. Several papers have discussed the flawed uses of the normal distribution model in meta-analyses [6-8], but to date only a few effective methods have been developed to address this issue, and there are no useful statistical packages that can be handled by non-statisticians. Also, there is a lack of real-world evidence that clearly demonstrates the relevance of this issue.

In this article, we propose random-effects meta-analysis methods with flexible distribution models that can flexibly express skewness, kurtosis, and tailweight: (1) skew normal distribution [9, 10], (2) skew *t*-distribution [10, 11], (3) asymmetric Subbotin distribution [10, 12], (4) Jones–Faddy distribution [13], and (5) sinh–arcsinh distribution [14]. Via application of these five flexible random-effects distribution models to two recently published systematic reviews [15, 16], we will demonstrate that the overall conclusions and interpretations of meta-analyses can be dramatically altered if the normal distribution assumption is not suitable. In addition, we provide a new R package, **flexmeta**, that can perform meta-analysis with simple code using the flexible random-effects distributions. We will explicitly show that the implicit uses of the normal distribution assumption might yield misleading results, and that our flexible alternative distributions may provide more valid conclusions for health technology assessments and policy making.

## Methods

### *Descriptions of two motivating meta-analyses*

We searched for recently published systematic reviews in leading medical journals (e.g., *BMJ*, *JAMA*), and found two examples [15, 16] that clearly demonstrated the unsuitability of the normal distribution assumption. The first example is a meta-analysis by Rubinstein et al. [15] assessing the benefits and harms of spinal manipulative therapy (SMT) for the treatment of chronic lower back pain. In Figure 1(a), we present a forest plot of their meta-analysis of 23 randomized controlled trials assessing the reduction of pain at 1 month (0–100; 0 = no pain, 100 = maximum pain) for SMT (N = 1629) vs. recommended therapies (N = 1526). The effect measure was the mean difference (MD). Using the ordinary random-effects meta-analysis method based on the normal random-effects distribution, we identified a substantial heterogeneity of the treatment effects, $I^2 = 92\%$, $\tau^2 = 112.20$ ($P < 0.01$; Cochrane's Q-test). The between-studies heterogeneity should be addressed in synthesis analysis. However, most of the MD estimates fell within relatively narrow range around the mean, although a small number exhibited larger effect sizes. This might imply that the true MD distribution is a skewed, heavy tailed, and sharp distribution. Although the average MD was estimated as −3.17 (95%CI: −7.85, 1.51) by the DerSimonian-Laird method [3], the point and interval estimates depend on the normal distribution assumption.

The second example is a meta-analysis by Koutoukidis et al. [16] aimed at estimating the association of weight loss interventions with biomarkers of liver disease in nonalcoholic fatty liver disease. In Figure 1(b), we also present a forest plot of their meta-analysis of 25 randomized controlled trials that assess the weight loss (kg) for more-intensive weight loss interventions (N = 1496) vs. no or lower-intensity weight loss interventions (N = 1062). Again, the effect measure was the MD, and we identified a substantial heterogeneity of the treatment effects: $I^2$ = 95%, $\tau^2$ = 12.45 (P < 0.01; Cochrane's Q-test). In this case, the MD estimates were not symmetrically distributed, and a certain number of trials exhibited a larger intervention effect than the average MD of −3.51 (95%CI: −5.03, −2.00). Thus, the true MD distribution would be a skewed, heavy tailed distribution. In particular, in predicting the intervention effect of a future trial, the normal distribution model would not suitably fit this dataset. Although the ordinary 95% Higgins–Thompson–Spiegelhalter (HTS) prediction interval [5] was (−11.02, 3.99), it might not express the true nature of the intervention effects in the target population.

*The flexible random-effects distribution models*

To address the restriction problem of the normal distribution, we propose random-effects meta-analysis methods using five flexible random-effects distribution models. For the notation, we consider that there are $K$ studies to be synthesized, and that $Y_i$ $(i = 1,2,\dots,K)$ is the estimated treatment effect measure in the $i$th study, e.g., mean difference, odds ratio, and hazard ratio; the ratio measures are typically transformed to logarithmic scales. The random-effects models considered here [2, 3] are then defined as

$$Y_i \sim N(\theta_i, \sigma_i^2) \qquad (*)$$

$$\theta_i \sim F(\theta)$$

where $\theta_i$ is the true effect size of the $i$th study, and $\sigma_i^2$ is the within-studies variance,

which is usually assumed to be known and fixed to valid estimates. Also, $F(\theta)$ corresponds to the random-effects distribution that expresses the heterogeneous probability distribution of $\theta_i$. For the conventional normal-normal random-effects model, $F(\theta)$ corresponds to a normal distribution. The predictive interval for future study [4] is substantially constructed based on the random-effects distribution $F(\theta)$. To overcome the limitations on the expressive ability of the normal distribution, our proposal is to adopt alternative flexible probability distributions.

Currently, due to developments in statistical distribution theory, various flexible probability distributions are available [10]. Here, we chose five representatives of the newest recently developed distributions. In Figure 2, we present some examples of these five distributions and their ability to express various shapes. Their mathematical details are presented in Supplementary data; here, we describe them in a non-technical manner.

*Skew normal distribution* $\mathrm{SN}(\xi, \omega, \alpha)$

The skew normal distribution [9, 10] is a generalized version of the conventional normal distribution that allows for skewness. $\xi$ is the location parameter that regulates the center location, and $\omega$ is the scale parameter that regulates the dispersion of the distribution; we use these notations similarly for the following four distributions as well; in the graphical displays in Figure 2, we set these parameters to $\xi = 0$ and $\omega = 1$, consistently. Also, $\alpha$ is the skewness parameter that adjusts the skewness, and the distribution is positively (negatively) skewed for $\alpha > 0$ ($\alpha < 0$). When $\alpha = 0$, it accords to a normal distribution $\mathrm{N}(\xi, \omega)$. In Figure 2(a), we present probability density functions of the skew normal distribution with $\alpha = 0,1,2,4,6,8$. It can flexibly express skew-shaped distributions, but has limitations in expressing kurtosis and tailweight.

*Skew t-distribution* $\mathrm{ST}(\xi, \omega, \nu, \alpha)$

The skew $t$-distribution [11] is also a generalized version of the conventional Student $t$-distribution that allows for skewness. The $t$-distribution can express heavy tailweight and kurtosis relative to the normal distribution via varying the degree of freedom $\nu$ (> 0). In this case, $\alpha$ is the skewness parameter; the distribution is positively (negatively) skewed for $\alpha > 0$ ($\alpha < 0$). Also, for $\alpha = 0$, it accords to the ordinary $t$-distribution. In Figure 2(b)(c), we present the skew $t$-distribution with $\alpha = 1,2,4,5,10,20$ with $\nu = 2,8$. The skew $t$-distribution can express flexible shapes by controlling the degree of freedom, relative to the skew normal distribution, especially for kurtosis and tailweight.

*Asymmetric Subbotin distribution (Type II)* AS2$(\xi, \omega, \nu, \alpha)$

Subbotin [17] proposed a symmetric probability distribution that can flexibly regulate the kurtosis and tail thickness; it involves a double exponential and trapezoidal-shaped distributions as special cases. An extended version of this distribution is the asymmetric Subbotin distribution of type II (AS2) [10, 12]. We present some examples of the AS2 distribution in Figure 2(d)(e), which clearly display its flexible expression ability. $\alpha$ is the skewness parameter, and $\nu$ (> 0) is the degree of freedom that regulates the kurtosis. The distribution is positively (negatively) skewed for $\alpha > 0$ ($\alpha < 0$), and more kurtosed for smaller $\nu$.

*Jones–Faddy distribution* JF$(\xi, \omega, a, b)$

Jones and Faddy [13] proposed another skewed version of the $t$-distribution. The Jones–Faddy distribution regulates the skewness and kurtosis by two parameters, $a$ (>0) and $b$ (>0); $\xi$ and $\omega$ are the location and scale parameters. Some examples are provided Figure 2(f)(g), some examples are provided. This distribution is positively (negatively) skewed for $a > b$ ($a < b$). Also, it reduces to $t$-distribution for $a = b$, with $a + b$ degrees of freedom. It can also flexible express various skewed, sharp, and heavy-tailed distributions by regulating the four parameters.

*Sinh–arcsinh distribution* SAS$(\xi, \omega, \delta, \epsilon)$

Jones and Pewsey [14] proposed a flexible unimodal four-parameter distribution that is induced by sinh–arcsinh (SAS) transformation. The SAS distribution can express symmetric and skewed shapes with heavy and light tailweight. In Figure 2(h)(i), several examples are presented. $\xi$ and $\omega$ are the location and scale parameters, $\delta$ $(> 0)$ is the kurtosis parameter, and $\epsilon$ is the skewness parameter. This distribution is positively (negatively) skewed for $\epsilon > 0$ $(\epsilon < 0)$. The kurtosis is regulated by $\delta$. It can express various skew $t$-distributions with quite sharp and gently sloped shapes with various degrees of skewness.

*Methods for the treatment effect estimation and prediction*

For the random-effects model (*), we can adopt the flexible distribution models for the random-effects distribution $F(\theta)$. The average treatment effect can be addressed as the mean $\mu$ of $F(\theta)$. As in the conventional DerSimonian-Laird-type normal-normal model, the parameters of $F(\theta)$ can be estimated by frequentist methods (e.g., the maximum likelihood estimation), but in many cases, they require complex numerical integrations; the computations of confidence intervals and P-values also have computational difficulties. Besides, through Bayesian approaches, we can compute posterior distributions of the mean parameter using a unified Markov Chain Monte Carlo (MCMC) framework[18, 19]. In addition, under the Bayesian framework, we can directly assess the predictive distribution of the treatment effect for a future study by the posterior predictive distribution[5]. When using the flexible parametric distributions, we can directly assess the nature and degree of heterogeneity by the predictive distribution. The variance of these distributions can be similarly defined for these distributions, but might not be properly interpreted as a dispersion parameter for skewed distributions. For these flexible skewed distributions, the predictive distributions can be directly used as a heterogeneity measure. Also, if we assume a non-informative prior distribution, the posterior inference can be substantially equivalent to the frequentist inference. For comparisons of competing models, we can use model assessment criteria of Bayesian statistics, e.g., the deviation information criterion (DIC) [20]. These computations can be easily performed by simple commands using the R package **flexmeta** (available at https://github.com/nomahi/flexmeta). The methodological details for the Bayesian modelling are presented in Supplementary data.

*Applications to the two meta-analyses*

We applied the flexible random-effects models to the two meta-analysis datasets described above. As reference methods, we also conducted the same analyses using the normal and $t$-distribution models. We used R ver. 3.5.1 (R Foundation for Statistical Computing, Vienna, Austria) and the **flexmeta** package for the statistical analyses; to implement MCMC, we used RStan ver. 2.19.2 [21]. After 10000 warm-ups, 250000 samples were used for the posterior inferences and prediction. The 95% credible intervals (CrI) and predictive intervals (PI) were calculated using the posterior samples of the mean of $F(\theta)$ and the predictive distribution of the effect of a future study $\theta_{new} \sim F(\theta)$ from MCMC. To evaluate the impacts of adopting the flexible distribution models rather than the ordinary normal distribution, we present graphical displays of the posterior and predictive distributions. In addition, we assessed model adequacies by DIC.

## Results

In Table 1 (a), we present the summary of the posterior distributions for the mean $\mu$ of the random-effects distributions. For the first example, the meta-analysis of chronic lower back pain, the posterior summary of the normal distribution is similar to the results of the conventional method, and the overall MD is −3.17 (95%CrI: −8.02, 1.73). The posterior means and 95%CrI of $\mu$ were quite different. In Figure 3, we present graphical displays of the 250000 posterior samples of $\mu$. All of the estimated posterior distributions by the five flexible random-effects distribution models indicated skewed and sharp-shaped distributions; it was sharper even for the $t$-distribution. DIC comparisons suggested that the best-fitting model was the AS2 distribution (DIC = 139.03); the SAS distribution was comparable to it (DIC = 139.97). Both of these distributions yielded larger MD estimates: −3.99 (95%CrI: −9.47, −0.10) and −5.33 (95%CrI: −11.37, −0.94), respectively. In addition, the posterior probabilities that $\mu$ is smaller than 0 were 0.98 and 0.99, respectively, whereas that of the normal random-effects distribution model was 0.90. In the original paper by Rubinstein et al. [15], the overall MD test was not statistically significant at the 5% level. However, the overall results were clearly altered by adopting the skewed flexible distribution models, which strongly indicated that the true effect sizes would lie within a narrower range and would be skew distributed. The overall conclusion for the overall MD could be changed using the flexible models. Also, we present a summary of the predictive distribution of this example in Table 2 (a). These results also indicated that the predictive distribution would be strongly skewed.

For the second example, the meta-analysis of non-alcoholic fatty liver disease, we present summaries of the posterior distribution of $\mu$ and the predictive distribution in Table 1(b) and 2(b). For this case, the normal distribution model provides results that are similar to those of the conventional methods. However, DIC comparisons revealed that the normal distribution was the worst-fitted model (DIC = 99.27), whereas AS2 and skew $t$-distribution were the best-fitted (DIC = 94.72, 95.44). For the overall MD $\mu$, the flexible models exhibited more skewed posterior distributions and larger MD estimates. Further, in Figure 4, we present the predictive distributions of the seven distribution models. We found that all of the flexible distribution models exhibited skewed and sharp-shaped distributions. In particular, the well-fitted AS2 and skew $t$-distributions indicated that the treatment effect in a future study $\theta_{new}$ would lie within a narrower range and would be skew distributed. However, the 95%PI of the normal distribution model was (−10.71, 3.68), those of AS2 and skew $t$-distribution were (−12.96, 0.15) and (−11.53, 0.55), respectively. The posterior probabilities of $\Pr(\theta_{new} < 0)$ of the normal distribution model was 0.84, whereas those of AS2 and skew $t$-distribution were 0.97 and 0.95, respectively. Hence, the overall conclusions might be altered.

## Discussion

Conclusions obtained from meta-analyses are widely applied to public health, clinical practice, health technology assessments, and policy-making. If misleading results have been produced by inadequate methods, the impact might be enormous. In this article, we proposed effective methods for meta-analysis using flexible random-effects distribution models, and provided an easily implementable statistical package for these methods. Through illustrative examples, we clearly showed the restrictions of using the conventional normal random-effects distribution model, which may yield misleading conclusions. The flexible random-effects distribution models represent effective tools for preventing such an outcome. Conventionally, these MCMC computations require special software and high-performance computers; to address these obstacles, we developed a user-friendly package, **flexmeta**, which was designed to be easily handled by non-statisticians and is freely available online. The proposed methods and the developed tools would help us to provide precise evidence. At a minimum, we recommend using these methods in sensitivity analyses.

In this study, we adopted five flexible distributions, but other probability distributions exist in statistical theory, e.g., see the comprehensive textbook by Azzalini and Capitanio [10]. Other choices might also be considered, but the five distribution models discussed here have sufficient expressive abilities, and significantly different results would not be obtained by adopting other existing distributions. Another choice would be to adopt nonparametric methods [18, 19]. However, in meta-analysis in medical studies, the number of studies $K$ is usually not large [22, 23]; consequently, nonparametric methods would be unstable in many applications because they require much larger statistical information (parallel to $K$) to conduct valid estimation and prediction.

As shown by the real data applications, existing meta-analyses may have reached misleading conclusions due to the straightforward uses of the normal random-effects distribution model. Our proposed methods might change the overall conclusions of these meta-analyses, and systematic re-evaluation of existing meta-analyses would be an interesting topic for future studies. In addition, for future systematic reviews, the flexible methods might be used as standard methods to provide accurate conclusions, at least in sensitivity analyses.

## Acknowledgements

This study was supported by Grant-in-Aid for Scientific Research from the Japan Society for the Promotion of Science (Grant numbers: JP19H04074, JP17K19808).

**Table 1**. Summary of the posterior distributions for the mean $\mu$ of the random-effects distribution [15, 16 †].

| Random-effects distribution | Mean | SD | 95% CrI | Pr($\mu$ < 0) | DIC |
|---|---|---|---|---|---|
| (a) Meta-analysis of the treatment of chronic low back pain | | | | | |
| Normal distribution | −3.17 | 2.47 | (−8.02, 1.73) | 0.90 | 145.62 |
| *t*-distribution | −1.43 | 1.91 | (−5.47, 2.14) | 0.78 | 143.54 |
| Skew normal distribution | −4.27 | 2.05 | (−8.37, −0.32) | 0.98 | 141.91 |
| Skew *t*-distribution | −3.47 | 2.09 | (−8.04, 0.19) | 0.97 | 140.50 |
| AS2 distribution | −3.99 | 2.39 | (−9.47, −0.10) | 0.98 | 139.03 |
| Jones–Faddy distribution | −3.09 | 2.05 | (−7.49, 0.65) | 0.95 | 141.94 |
| Sinh–arcsinh distribution | −5.33 | 2.91 | (−11.37, −0.94) | 0.99 | 139.97 |
| (b) Meta-analysis of the treatment of nonalcoholic fatty liver disease | | | | | |
| Normal distribution | −3.52 | 0.77 | (−5.04, −1.98) | 1.00 | 99.27 |
| *t*-distribution | −3.06 | 0.66 | (−4.43, −1.81) | 1.00 | 97.95 |
| Skew normal distribution | −3.83 | 0.69 | (−5.29, −2.55) | 1.00 | 96.52 |
| Skew *t*-distribution | −3.61 | 0.69 | (−5.14, −2.40) | 1.00 | 95.44 |
| AS2 distribution | −3.78 | 0.83 | (−5.72, −2.44) | 1.00 | 94.72 |
| Jones–Faddy distribution | −3.49 | 0.68 | (−4.93, −2.23) | 1.00 | 97.02 |
| Sinh–arcsinh distribution | −4.24 | 0.86 | (−6.03, −2.75) | 1.00 | 96.70 |

[†] CrI: credible interval, DIC: deviance information criterion.

**Table 2**. Summary of the predictive distributions for the two meta-analyses [15, 16 †].

| Random-effects distribution | Mean | SD | 95% PI | Pr($\theta_{new}$< 0) |
|---|---|---|---|---|
| (a) Meta-analysis for the treatment of chronic low back pain | | | | |
| Normal distribution | −3.17 | 11.33 | (−25.62, 19.29) | 0.61 |
| *t*-distribution | −1.43 | 9.69 | (−20.46, 17.35) | 0.58 |
| Skew normal distribution | −4.26 | 9.43 | (−26.42, 9.92) | 0.63 |
| Skew *t*-distribution | −3.47 | 9.60 | (−26.85, 8.51) | 0.59 |
| AS2 distribution | −4.22 | 10.56 | (−32.22, 6.81) | 0.58 |
| Jones−Faddy distribution | −3.09 | 9.65 | (−24.65, 11.22) | 0.59 |
| Sinh−arcsinh distribution | −5.37 | 11.80 | (−33.51, 7.67) | 0.60 |
| (b) Meta-analysis for the treatment of nonalcoholic fatty liver disease | | | | |
| Normal distribution | −3.52 | 3.63 | (−10.71, 3.68) | 0.84 |
| *t*-distribution | −3.06 | 3.37 | (−9.77, 3.51) | 0.86 |
| Skew normal distribution | −3.83 | 3.18 | (−11.31, 0.93) | 0.93 |
| Skew *t*-distribution | −3.61 | 3.23 | (−11.53, 0.55) | 0.95 |
| AS2 distribution | −3.81 | 3.57 | (−12.96, 0.15) | 0.97 |
| Jones−Faddy distribution | −3.49 | 3.20 | (−10.78, 1.72) | 0.91 |
| Sinh−arcsinh distribution | −4.31 | 3.72 | (−12.78, 0.61) | 0.95 |

[†] PI: predictive interval.

(a)

| Study | Experimental Total | Experimental Mean | Experimental SD | Control Total | Control Mean | Control SD | MD | 95%-CI | Weight (fixed) | Weight (random) |
|---|---|---|---|---|---|---|---|---|---|---|
| Krekoukiasa 2017 | 25 | 12.00 | 11.0000 | 25 | 49.60 | 8.9000 | -37.60 | [-43.15; -32.05] | 5.0% | 4.7% |
| Wilkey 2008 | 18 | 42.80 | 22.5000 | 12 | 70.00 | 24.1000 | -27.20 | [-44.35; -10.05] | 0.5% | 3.0% |
| Sarker 2017 | 35 | 20.30 | 6.2000 | 35 | 39.70 | 7.1000 | -19.40 | [-22.52; -16.28] | 15.6% | 5.0% |
| Skillgate 2007 | 92 | 36.00 | 14.4000 | 80 | 44.00 | 13.4000 | -8.00 | [-12.16; -3.84] | 8.8% | 4.9% |
| Gudavalli 2006 | 123 | 17.40 | 22.3000 | 112 | 23.40 | 20.7000 | -6.00 | [-11.50; -0.50] | 5.0% | 4.7% |
| Hondras 2009 | 83 | 27.63 | 19.3100 | 16 | 33.47 | 19.4900 | -5.84 | [-16.25; 4.57] | 1.4% | 4.1% |
| Hurwitz 2002 | 169 | 31.00 | 18.0000 | 168 | 35.00 | 20.0000 | -4.00 | [ -8.06; 0.06] | 9.2% | 4.9% |
| Hondras 2009 | 90 | 29.49 | 19.2900 | 16 | 33.47 | 19.4900 | -3.98 | [-14.33; 6.37] | 1.4% | 4.1% |
| Bronfort 1996 | 62 | 34.00 | 19.0000 | 43 | 36.00 | 22.0000 | -2.00 | [-10.10; 6.10] | 2.3% | 4.4% |
| Hurwitz 2002 | 169 | 34.00 | 19.0000 | 169 | 36.00 | 19.0000 | -2.00 | [ -6.05; 2.05] | 9.3% | 4.9% |
| Ulger 2017 | 57 | 20.80 | 12.0000 | 56 | 21.00 | 16.0000 | -0.20 | [ -5.42; 5.02] | 5.6% | 4.8% |
| UK BEAM trial 2004 | 316 | 48.99 | 22.1600 | 228 | 48.93 | 21.4900 | 0.06 | [ -3.65; 3.77] | 11.1% | 4.9% |
| Hemmila 2002 | 22 | 30.50 | 15.0000 | 35 | 30.00 | 15.0000 | 0.50 | [ -7.50; 8.50] | 2.4% | 4.4% |
| Dougherty 2014 | 32 | 37.50 | 28.9000 | 28 | 36.50 | 33.8000 | 1.00 | [-15.03; 17.03] | 0.6% | 3.2% |
| Bronfort 2011 | 50 | 39.00 | 18.0000 | 95 | 37.00 | 18.0000 | 2.00 | [ -4.16; 8.16] | 4.0% | 4.7% |
| Bronfort 2011 | 50 | 39.00 | 18.0000 | 96 | 36.00 | 20.0000 | 3.00 | [ -3.40; 9.40] | 3.7% | 4.6% |
| Cecchi 2010 | 35 | 20.00 | 20.0000 | 68 | 16.70 | 13.3000 | 3.30 | [ -4.04; 10.64] | 2.8% | 4.5% |
| Hemmila 2002 | 22 | 30.50 | 15.0000 | 34 | 27.00 | 15.0000 | 3.50 | [ -4.54; 11.54] | 2.4% | 4.4% |
| Rasmussen-Barr 2003 | 19 | 24.00 | 26.7000 | 22 | 20.00 | 17.8000 | 4.00 | [-10.12; 18.12] | 0.8% | 3.5% |
| Hsieh 2002 | 45 | 25.80 | 19.3000 | 42 | 21.30 | 12.8000 | 4.50 | [ -2.34; 11.34] | 3.3% | 4.6% |
| Cecchi 2010 | 35 | 20.00 | 20.0000 | 68 | 15.00 | 13.3000 | 5.00 | [ -2.34; 12.34] | 2.8% | 4.5% |
| Dougherty 2014 | 60 | 34.90 | 31.2000 | 61 | 26.60 | 33.5000 | 8.30 | [ -3.23; 19.83] | 1.1% | 3.9% |
| Waqqar 2016 | 20 | 25.50 | 26.5000 | 17 | 14.70 | 17.3000 | 10.80 | [ -3.43; 25.03] | 0.8% | 3.5% |
| **Fixed effect model** | **1629** | | | **1526** | | | **-5.91** | **[ -7.15; -4.68]** | **100.0%** | **--** |
| **Random effects model** | | | | | | | **-3.17** | **[ -7.85; 1.51]** | **--** | **100.0%** |

Heterogeneity: $I^2 = 92\%$, $\tau^2 = 112.2028$, $p < 0.01$

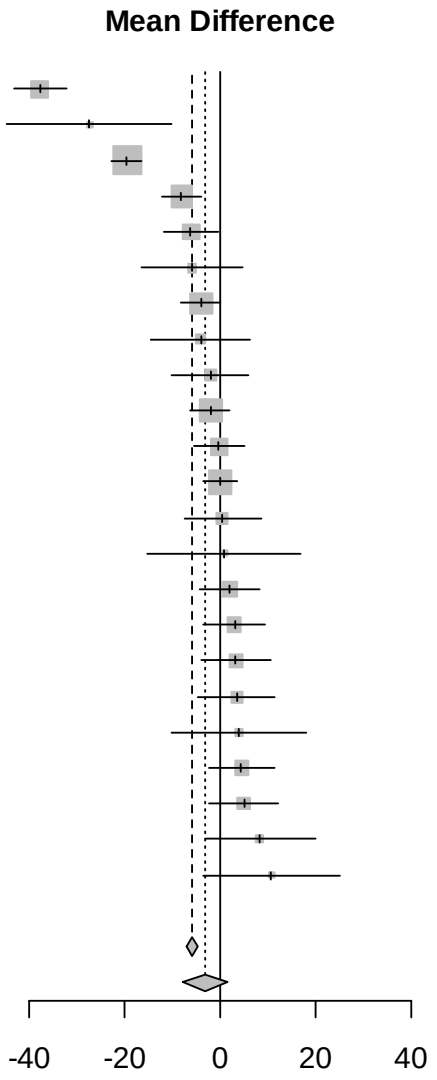

(b)

| Study | Experimental Total | Experimental Mean | Experimental SD | Control Total | Control Mean | Control SD | MD | 95%-CI | Weight (fixed) | Weight (random) |
|---|---|---|---|---|---|---|---|---|---|---|
| Al-Jiffrl et al, 2013 | 50 | -13.70 | 3.4000 | 50 | 0.80 | 3.0000 | -14.50 | [-15.76; -13.24] | 6.1% | 4.6% |
| Promrat et al, 2010 | 20 | -8.70 | 12.4000 | 10 | -0.50 | 7.4000 | -8.20 | [-15.31; -1.09] | 0.2% | 2.3% |
| Abd El-Kader et al, 2016 | 50 | -5.90 | 2.2000 | 50 | 2.10 | 2.9000 | -8.00 | [ -9.01; -6.99] | 9.4% | 4.7% |
| Sun et al, 2012 | 674 | -6.10 | 9.2000 | 332 | 1.30 | 9.2000 | -7.40 | [ -8.61; -6.19] | 6.5% | 4.6% |
| Bahmanadabi et al, 2011 | 20 | -13.00 | 10.0000 | 20 | -6.10 | 30.0000 | -6.90 | [-20.76; 6.96] | 0.0% | 1.0% |
| Abenavoli et al, 2017 | 20 | -5.10 | 3.4000 | 10 | 0.40 | 2.3000 | -5.50 | [ -7.56; -3.44] | 2.3% | 4.4% |
| Wong et al, 2013 | 77 | -5.60 | 5.7000 | 77 | -0.60 | 3.7000 | -5.00 | [ -6.52; -3.48] | 4.2% | 4.6% |
| Armstrong et al, 2016 | 23 | -5.30 | 4.7000 | 22 | -0.60 | 4.4000 | -4.70 | [ -7.36; -2.04] | 1.4% | 4.2% |
| Asghari et al, 2018 | 30 | -4.10 | 4.0000 | 30 | 0.00 | 2.7000 | -4.10 | [ -5.83; -2.37] | 3.2% | 4.5% |
| Axley et al, 2018 | 8 | -2.70 | 5.0000 | 14 | 0.90 | 6.0000 | -3.60 | [ -8.28; 1.08] | 0.4% | 3.3% |
| Selezneva et al, 2014 | 58 | -9.30 | 1.8000 | 116 | -6.20 | 1.7000 | -3.10 | [ -3.66; -2.54] | 30.8% | 4.7% |
| Katsagoni et al, 2018 (D) | 21 | -5.60 | 4.1000 | 11 | -2.80 | 3.9000 | -2.80 | [ -5.70; 0.10] | 1.1% | 4.1% |
| Lim et al, 2018 | 43 | -3.20 | 2.5000 | 43 | -0.90 | 2.0000 | -2.30 | [ -3.26; -1.34] | 10.5% | 4.7% |
| St George et al, 2009 (M) | 73 | -2.80 | 3.4000 | 17 | -0.50 | 3.7000 | -2.30 | [ -4.22; -0.38] | 2.6% | 4.4% |
| Katsagoni et al, 2018 (D+E) | 21 | -4.90 | 4.1000 | 10 | -2.80 | 3.9000 | -2.10 | [ -5.09; 0.89] | 1.1% | 4.0% |
| Zelber-Sagi et al, 2006 | 21 | -7.70 | 6.8000 | 23 | -5.90 | 5.9000 | -1.80 | [ -5.58; 1.98] | 0.7% | 3.7% |
| Harrison et al, 2009 | 23 | -8.20 | 5.0000 | 18 | -6.40 | 5.2000 | -1.80 | [ -4.95; 1.35] | 1.0% | 4.0% |
| Cheng et al, 2017 (D+E) | 29 | -1.50 | 3.2000 | 15 | 0.20 | 3.3000 | -1.70 | [ -3.74; 0.34] | 2.3% | 4.4% |
| Dong et al, 2016 | 130 | -1.40 | 4.9000 | 130 | 0.10 | 4.1000 | -1.50 | [ -2.60; -0.40] | 7.9% | 4.7% |
| St George et al, 2009 (L) | 36 | -1.90 | 3.7000 | 17 | -0.50 | 3.7000 | -1.40 | [ -3.53; 0.73] | 2.1% | 4.4% |
| Lee et al, 2012 | 8 | -1.60 | 3.1000 | 10 | -0.80 | 2.3000 | -0.80 | [ -3.38; 1.78] | 1.4% | 4.2% |
| Cheng et al, 2017 (D) | 28 | -0.80 | 3.1000 | 14 | -0.20 | 3.3000 | -0.60 | [ -2.68; 1.48] | 2.2% | 4.4% |
| Eckard et al, 2013 (MF) | 9 | -3.00 | 4.7000 | 5 | -2.50 | 5.3000 | -0.50 | [ -6.07; 5.07] | 0.3% | 2.9% |
| Khoo et al, 2017 | 12 | -3.50 | 2.1000 | 12 | -3.50 | 3.3000 | 0.00 | [ -2.21; 2.21] | 2.0% | 4.3% |
| Eckard et al, 2013 (LF) | 12 | -0.20 | 5.4000 | 6 | -2.50 | 5.3000 | 2.30 | [ -2.93; 7.53] | 0.4% | 3.0% |
| **Fixed effect model** | **1496** | | | **1062** | | | **-4.22** | **[ -4.53; -3.91]** | **100.0%** | **--** |
| **Random effects model** | | | | | | | **-3.51** | **[ -5.03; -2.00]** | **--** | **100.0%** |

Heterogeneity: $I^2 = 95\%$, $\tau^2 = 12.5494$, $p < 0.01$

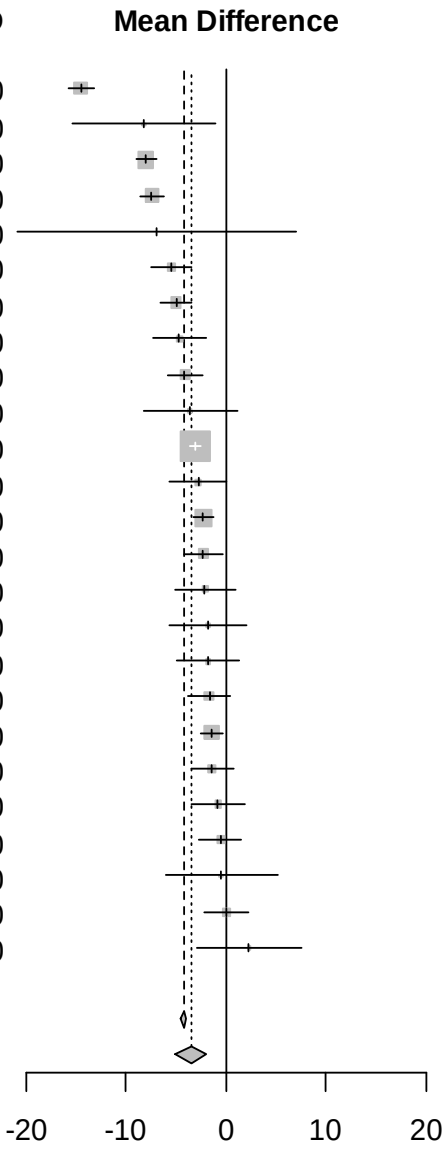

**Figure 1.** Forest plots for the two motivating examples: (a) meta-analysis of chronic low back pain [15], (b) meta-analysis of nonalcoholic fatty liver disease [16].

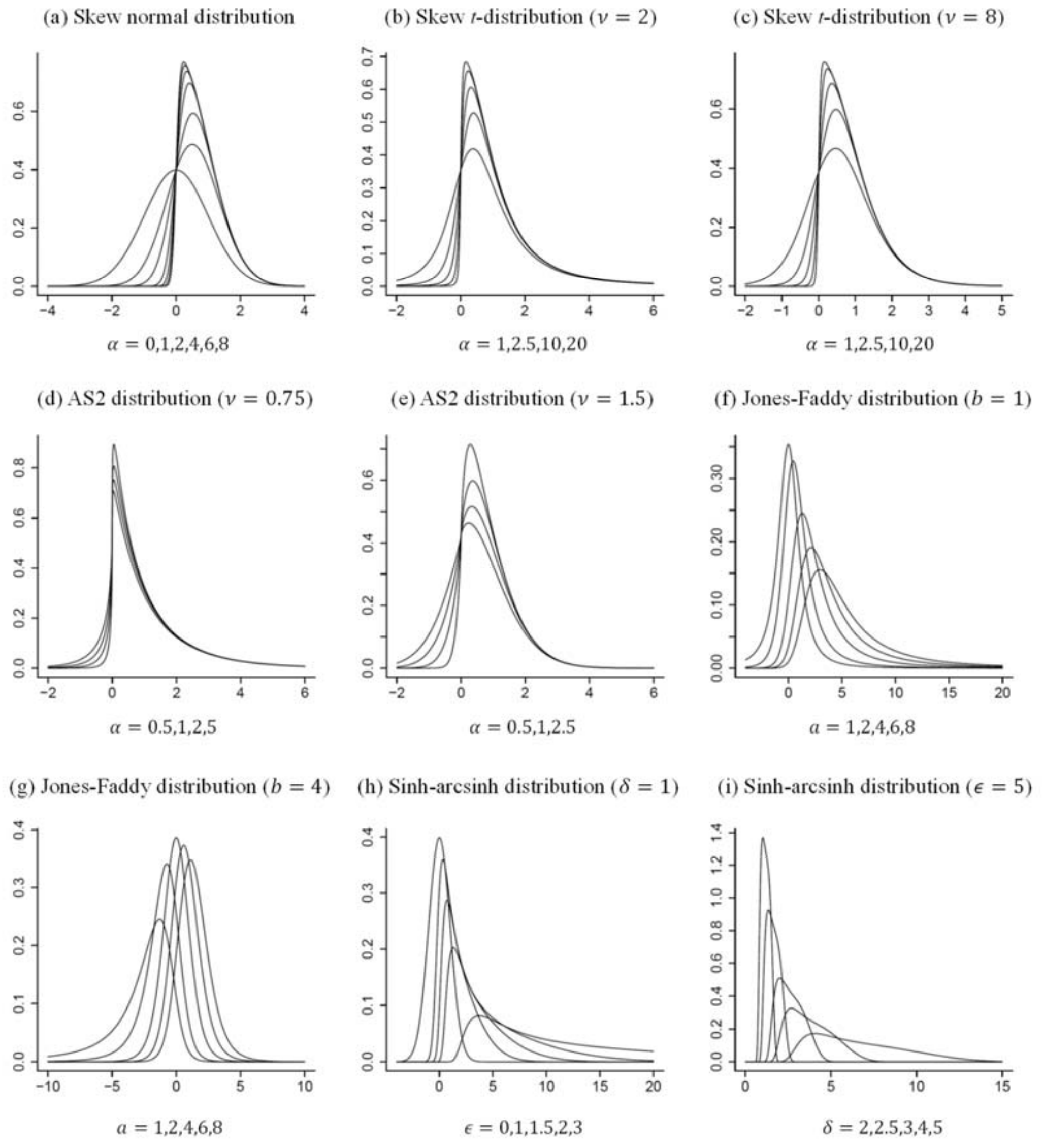

**Figure 2.** Flexible models for the random-effects distribution: (a) skew normal distribution, (b), (c) skew *t*-distribution, (d), (e) AS2 distribution, (f), (g) Jones–Faddy distribution, (h), (i) sinh–arcsinh distribution.

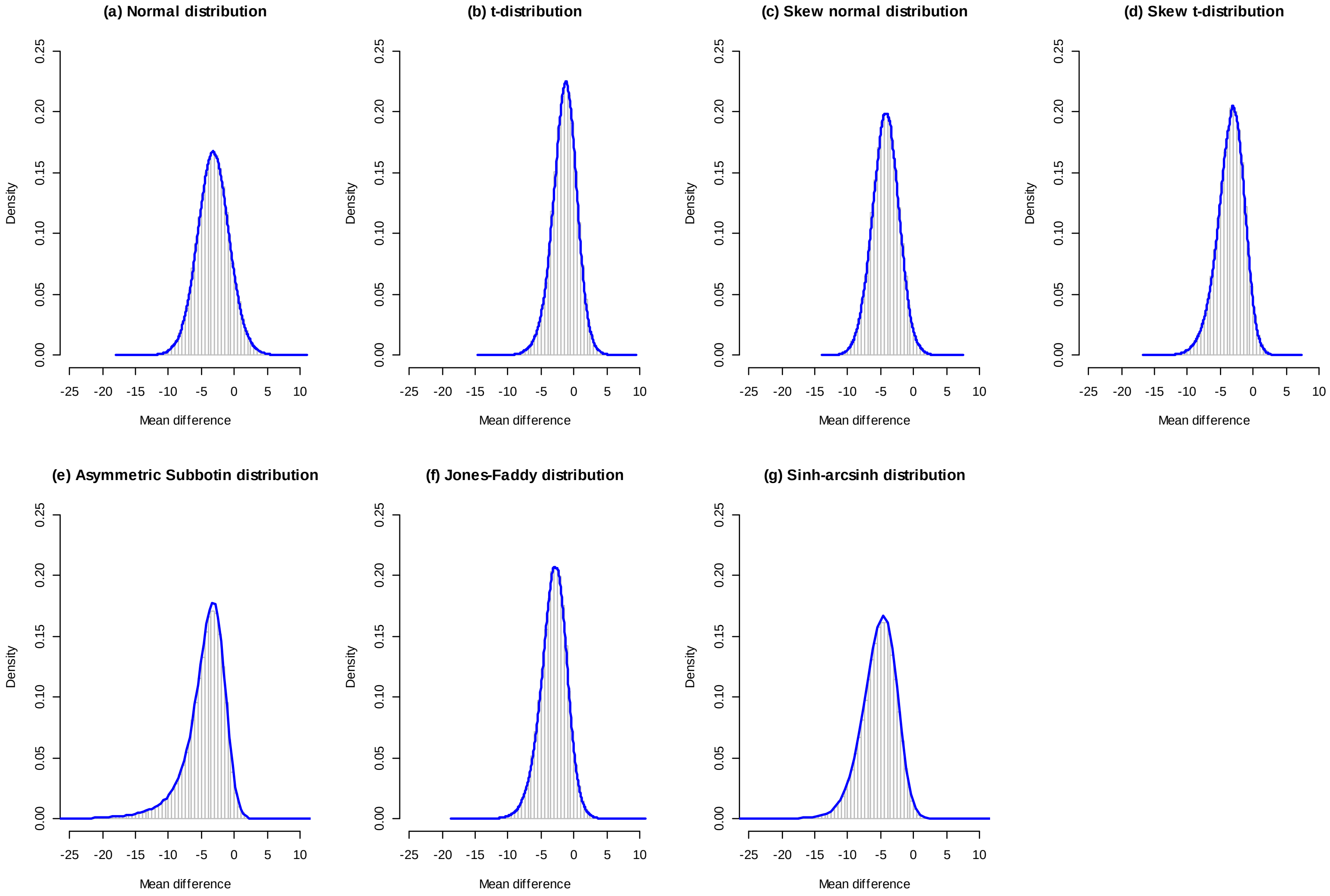

**Figure 3.** Posterior distributions for $\mu$ of the meta-analysis of chronic low back pain [15] using seven random-effects distribution models.

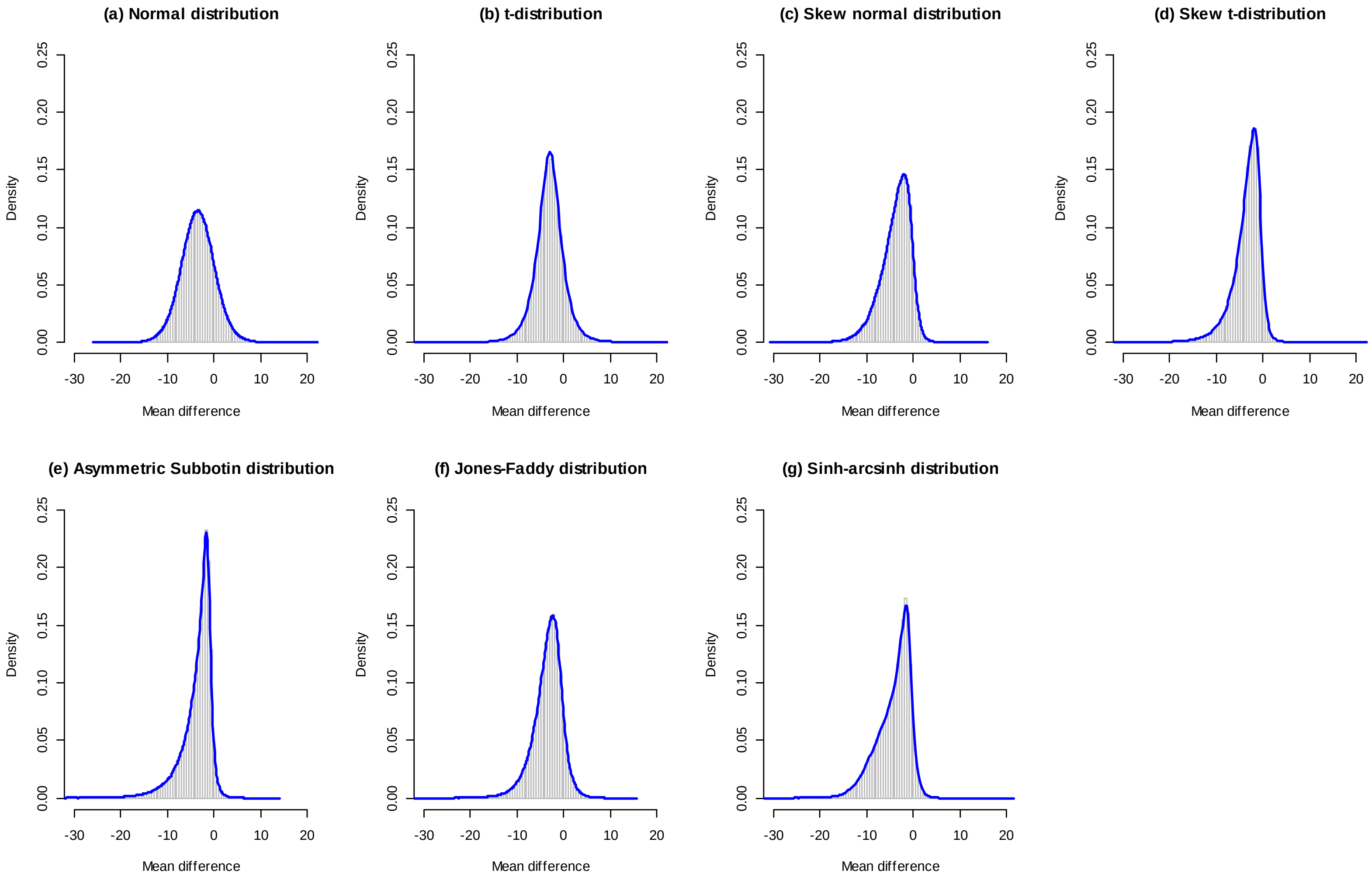

**Figure 4.** Predictive distributions for the meta-analysis of nonalcoholic fatty liver disease [16] using seven random-effects distribution models.

SUPPLEMENTARY DATA for

# Meta-analysis using flexible random-effects distribution models

Hisashi Noma[1], Kengo Nagashima[2], Shogo Kato[3], Satoshi Teramukai[4]
and Furukawa, T. A.[5]

[1] *Department of Data Science, The Institute of Statistical Mathematics, Tokyo, Japan*
[2] *Research Center for Medical and Health Data Science, The Institute of Statistical Mathematics, Tokyo, Japan*
[3] *Department of Statistical Inference and Mathematics, The Institute of Statistical Mathematics, Tokyo, Japan*
[4] *Department of Biostatistics, Graduate School of Medical Science, Kyoto Prefectural University of Medicine, Kyoto, Japan*
[5] *Departments of Health Promotion and Human Behavior, Kyoto University Graduate School of Medicine/School of Public Health, Kyoto, Japan*

## e-Appendix A: Mathematical details of the flexible distributions

### *A.1 Skew normal distribution*

The skew normal distribution (Azzalini, 1985; Azzalini and Capitanio, 2014) is a generalized version of the conventional normal distribution that allows for skewness. The probability density function is

$$p(\theta|\xi,\omega,\alpha) = \frac{2}{\omega}\phi\left(\frac{\theta-\xi}{\omega}\right)\Phi\left(\frac{\alpha(\theta-\xi)}{\omega}\right)$$

where $\phi(\theta)$ and $\Phi(\theta)$ are, respectively, the probability density and cumulative distribution functions of the standard normal distribution N(0, 1). $\xi$ is the location parameter that regulates the center location, and $\omega$ (> 0) is the scale parameter that regulates the dispersion of the distribution. $\alpha$ is the skewness parameter that adjusts the skewness; the distribution is positively (negatively) skewed for $\alpha > 0$ ($\alpha < 0$). When $\alpha = 0$, the skew normal distribution accords with the normal distribution $\mathrm{N}(\xi,\omega)$. The mean and variance of this distribution are

$$\mathrm{E}[\theta] = \xi + \omega b\delta$$

$$\mathrm{V}[\theta] = \omega^2[1-(b\delta)^2]$$

where

$$b = \sqrt{\frac{2}{\pi}}$$

$$\delta = \frac{\alpha}{\sqrt{1+\alpha^2}}$$

*A.2 Skew t-distribution*

The skew *t*-distribution (Azzalini and Capitanio, 2003, 2014) is also a generalized version of the conventional Student *t*-distribution that allows for skewness. The probability density function is

$$p(\theta|\xi,\omega,\nu,\alpha) = \frac{2}{\omega} f_t\left(\frac{\theta-\xi}{\omega}\middle|\nu\right) F_t\left(\frac{\alpha(\theta-\xi)}{\omega}\sqrt{\frac{\omega^2(\nu+1)}{\nu\omega^2+(\theta-\xi)^2}}\middle|\nu+1\right)$$

where $f_t(\theta|\nu)$ and $F_t(\theta|\nu)$ are, respectively, the probability density and cumulative distribution functions of the Student *t*-distribution with $\nu$ (> 0) degrees of freedom. $\xi$ is the location parameter that regulates the center location, and $\omega$ (> 0) is the scale parameter that regulates the dispersion of the distribution. $\alpha$ is the skewness parameter that adjusts the skewness and the distribution is positively (negatively) skewed for $\alpha > 0$ $(\alpha < 0)$. When $\alpha = 0$, this distribution accords with the Student *t*-distribution with $\nu$ degrees of freedom. The mean and variance of this distribution are

$$\mathrm{E}[\theta] = \xi + \omega b_\nu \delta$$
$$\mathrm{V}[\theta] = \omega^2\left[\frac{\nu}{\nu-2} - (b_\nu\delta)^2\right]$$

where

$$b_\nu = \frac{\sqrt{\nu}\Gamma((\nu-1)/2)}{\sqrt{\pi}\Gamma(\nu/2)}$$

$$\delta = \frac{\alpha}{\sqrt{1+\alpha^2}}$$

The skew *t*-distribution can express flexible shapes by controlling the degree of freedom,

compared with skew normal distribution, especially for the kurtosis and tailweight.

*A.3 Asymmetric Subbotin distribution*

Subbotin (1923) proposed a symmetric probability distribution that can regulate the kurtosis and tail thickness flexibly. The probability density function of the Subbotin distribution with $\nu$ (> 0) degrees of freedom is

$$f_S(\theta) = \frac{1}{2\nu^{1/\nu}\Gamma(1+1/\nu)} \exp\left(-\frac{|\theta|^\nu}{\nu}\right)$$

Based on the form of the probability density function, this distribution involves a double exponential and trapezoidal-shaped distributions as special cases. The location and scale can be regulated by linear transmission, and the distribution can flexibly express heavy and light tailweight. The asymmetric Subbotin distribution of type II (AS2) (Azzalini, 1986) is an extended version of this distribution, and the probability density function is

$$p(\theta|\xi,\omega,\nu,\alpha) = \frac{2}{\omega} f_S\left(\frac{\theta-\xi}{\omega}\middle|\nu\right) F_S\left(\frac{\alpha(\theta-\xi)}{\omega}\right)$$

where

$$F_S(\theta) = \Phi\left(\operatorname{sgn}(\theta)\frac{|\theta|^{\nu/2}}{\sqrt{\nu/2}}\right)$$

$\xi$ is the location parameter that regulates the center location, and $\omega$ is the scale parameter that regulates the dispersion of the distribution. The distribution is positively (negatively) skewed for $\alpha > 0$ ($\alpha < 0$), and more kurtosed for smaller $\nu$. The mean and variance of this distribution are

$$\mathrm{E}[\theta] = \xi + \operatorname{sgn}(\alpha)\omega C_\nu Q_\nu$$

$$\mathrm{V}[\theta] = \omega^2\left[\frac{\nu^{2/\nu}\Gamma(3/\nu)}{\Gamma(1/\nu)} - (C_\nu Q_\nu)^2\right]$$

where

$$C_\nu = \frac{\nu^{1/\nu}\Gamma(2/\nu)}{\Gamma(1/\nu)}$$

$$Q_\nu = 2F_t\left(\sqrt{4|\alpha|^\nu/\nu}\,\middle|\, 4/\nu\right) - 1$$

As shown in Figure 2, the AS2 distribution can express a sharp skew distribution, which can be seen as an asymmetric double exponential distribution. Also, it can express a more rounded shape, like the skew *t*-distribution.

### *A.4 Jones-Faddy distribution*

Jones and Faddy (2003) proposed another skewed version of *t*-distribution, whose probability density function is expressed as

$$p(\theta|\xi,\omega,\nu,\alpha) = \frac{1}{\omega} f_{JF}\left(\frac{\theta-\xi}{\omega}\,\middle|\, a,b\right)$$

where

$$f_{JF}(z|a,b) = C_{a,b}^{-1}\left\{1+\frac{z}{(a+b+z^2)^{1/2}}\right\}^{a+1/2}\left\{1-\frac{z}{(a+b+z^2)^{1/2}}\right\}^{b+1/2}$$

$$C_{a,b} = 2^{a+b-1}B(a,b)(a+b)^{1/2}$$

The Jones–Faddy distribution regulates the skewness and kurtosis through two model parameters $a$ $(> 0)$ and $b$ $(> 0)$. $\xi$ is the location parameter that regulates the center location, and $\omega$ (> 0) is the scale parameter. This distribution is positively (negatively) skewed for $a > b$ ($a < b$). Also, it reduces to the *t*-distribution for $a = b$, with $a + b$ degrees of freedom. It can also flexibly express various distributions involving skewed, sharp, and heavy-tailed shapes by regulating the four parameters. The mean and variance are

$$\mathrm{E}[\theta] = \xi + \omega\eta_{a,b}$$

$$\mathrm{V}[\theta] = \omega^2\left[\frac{a+b}{4}\frac{(a-b)^2+a+b-2}{(a-1)(b-1)} - {\eta_{a,b}}^2\right]$$

where

$$\eta_{a,b} = \frac{(a-b)\sqrt{a+b}}{2}\frac{\Gamma(a-1/2)\Gamma(b-1/2)}{\Gamma(a)\Gamma(b)}$$

*A.5 Sinh–arcsinh distribution*

Jones and Pewsey (2009) proposed a flexible unimodal four parameter distribution that is induced by sinh–arcsinh (SAS) transformation. The probability density function is

$$p(\theta|\xi,\omega,\epsilon,\delta) = \frac{1}{\omega} f_{SAS}\left(\frac{\theta-\xi}{\omega}\middle|\epsilon,\delta\right)$$

where

$$f_{SAS}(z|\epsilon,\delta) = \frac{1}{\sqrt{2\pi(1+z^2)}}\delta C_{\epsilon,\delta}(z)\exp\left(-\frac{S_{\epsilon,\delta}^2(z)}{2}\right)$$

$$C_{\epsilon,\delta}(z) = \cosh\{\delta\sinh^{-1}(z)-\epsilon\}$$

$$S_{\epsilon,\delta}(z) = \sinh\{\delta\sinh^{-1}(z)-\epsilon\}$$

The SAS distribution can express symmetric and skewed shapes with heavy and light tailweight. $\xi$ and $\omega$ are the location and scale parameters, $\delta$ $(>0)$ is the kurtosis parameter, and $\epsilon$ is the skewness parameter. This distribution is positively (negatively) skewed for $\epsilon > 0$ $(\epsilon < 0)$. The kurtosis is regulated by $\delta$. It can express different shapes with the skew *t*-distributions involving quite sharp and gently sloped ones with various degrees of skewness. The mean and variance of this distribution are

$$\mathrm{E}[\theta] = \xi + \omega\zeta_{\delta,\epsilon}$$

$$\mathrm{V}[\theta] = \omega^2\left[\lambda_{\delta,\epsilon} - \zeta_{\delta,\epsilon}^2\right]$$

where

$$\zeta_{\delta,\epsilon} = \frac{e^{1/4}}{\sqrt{8\pi}}\sinh\left(\frac{\epsilon}{\delta}\right)\left\{K_{(1/\delta+1)/2}\left(\frac{1}{4}\right) + K_{(1/\delta-1)/2}\left(\frac{1}{4}\right)\right\}$$

$$\lambda_{\delta,\epsilon} = \frac{1}{2}\left\{\frac{e^{1/4}}{\sqrt{8\pi}}\cosh\left(\frac{2\epsilon}{\delta}\right)\left\{K_{(2/\delta+1)/2}\left(\frac{1}{4}\right) + K_{(2/\delta-1)/2}\left(\frac{1}{4}\right)\right\} - 1\right\}$$

and $K_a(z)$ is the modified Bessel function of the second kind.

## e-Appendix B: Methods for the Bayesian modelling

For the Bayesian random-effects model,

$$Y_i \sim N(\theta_i, \sigma_i^2) \qquad (*)$$

$$\theta_i \sim F(\theta)$$

the prior distributions for the model parameters of $F(\theta)$ directly influence to the posterior inferences and predictions. We adopted non-informative priors in the applications, and the default settings of **flexmeta** also adopt the same prior distributions. Here, we explain which priors were adopted; the source R and Stan codes are available at our GitHub site (https://github.com/nomahi/flexmeta), and the users can freely customize the prior distribution settings.

For the location and scale parameters $\xi$ and $\omega$, we consistently adopted the following vague prior distributions for the seven models (involving the ordinary normal and *t*-distribution models):

$$\xi \sim N(0, 100^2)$$

$$\omega \sim U(0, 20)$$

For the degree-of-freedom parameter $\nu$ of the *t*-distribution, skew *t*-distribution, and AS2 distribution, we adopted an exponential (0.1) prior that was restricted to $k > 2.5$ to assure the existence of the second moment ($k \geq 2$), as in Fernandez and Steel (1998) and Lee and Thompson (2008). For the skewness parameter $\alpha$ of the skew normal distribution, skew-*t* distribution, and AS2 distribution, we adopted a proper vague normal prior $N(0, 5^2)$. For the Jones–Faddy distribution, we assumed uniform priors for the two model parameters $a$ and , $a, b \sim U(1.5, 200)$. The lower bound of the uniform distribution is determined to assure the existences of the first, second and third moments (Jones and Faddy, 2003). For the SAS distribution, we also adopted vague priors for the skewness and kurtosis parameters, $\epsilon \sim N(0,100^2)$ and $\delta \sim U(0,100)$.